\documentclass{aastex}

\usepackage{spr-astr-addons}

\usepackage{graphicx,natbib}
\usepackage{amssymb}
\newcommand{\be}{\begin{equation}}
\newcommand{\ba}{\begin{eqnarray}}
\newcommand{\ee}{\end{equation}}
\newcommand{\ea}{\end{eqnarray}}

\newcommand{\gt}{>}
\def\lesssim{\mathrel{\hbox{\rlap{\hbox{\lower4pt\hbox{$\sim$}}}\hbox{$<$}}}}
\def\gtrsim{\mathrel{\hbox{\rlap{\hbox{\lower4pt\hbox{$\sim$}}}\hbox{$>$}}}}

\def\simless{\mathbin{\lower 3pt\hbox
   {$\rlap{\raise 5pt\hbox{$\char'074$}}\mathchar''7218$}}}   
\def\simgreat{\mathbin{\lower 3pt\hbox
   {$\rlap{\raise 5pt\hbox{$\char'076$}}\mathchar''7218$}}}   
\def\apj{ApJ}
\def\apjs{ApJS}
\def\apjl{ApJL}
\def\aap{A\&A}
\def\aj{AJ}
\def\mnras{MNRAS}

\def\physrep{{Phys. Reports}}
\shorttitle{Scales and Observability of Reionization}
\shortauthors{Iliev et al.}

\begin{document}


\title{Reionization: Characteristic Scales, Topology and Observability}

\author{Ilian T. Iliev}

\affil{Canadian Institute for Theoretical Astrophysics, University
  of Toronto, 60 St. George Street, Toronto, ON M5S 3H8, Canada}

\affil{Universit\"at Z\"urich, Institut f\"ur Theoretische Physik,
Winterthurerstrasse 190, CH-8057 Z\"urich, Switzerland}

\email{iliev@physik.uzh.ch}

\author{Paul R. Shapiro}
\affil{Department of Astronomy, University of Texas, Austin, TX 78712-1083,
  U.S.A.}

\author{Garrelt Mellema}
\affil{Stockholm Observatory, AlbaNova
  University Center, Stockholm University, SE-106 91 Stockholm, Sweden}

\author{Ue-Li Pen}
\author{Patrick McDonald}
\affil{Canadian Institute for Theoretical Astrophysics, University
  of Toronto, 60 St. George Street, Toronto, ON M5S 3H8, Canada}

\and
\author{Marcelo A. Alvarez}
\affil{Kavli Institute for Particle Astrophysics and Cosmology, Stanford
University, Stanford, CA 94305, USA}

\begin{abstract}
Recently the numerical simulations of the process of reionization of the
universe at $z>6$ have made a qualitative leap forward, reaching
sufficient sizes and dynamic range to determine the characteristic scales 
of this process. This allowed making the first realistic predictions for a
variety of observational signatures. We discuss recent results from
large-scale radiative transfer and structure formation simulations on the
observability of high-redshift Ly-$\alpha$ sources. We also briefly discuss
the dependence of the characteristic scales and topology of the ionized
and neutral patches on the reionization parameters. 
\end{abstract}




\begin{figure}
  \includegraphics[width=3.3in]{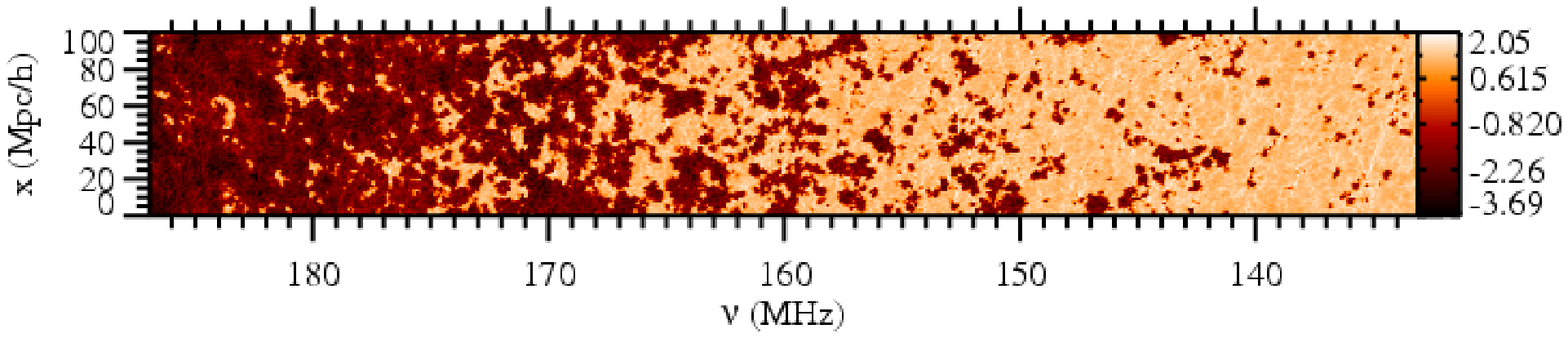}
  \includegraphics[width=3.2in]{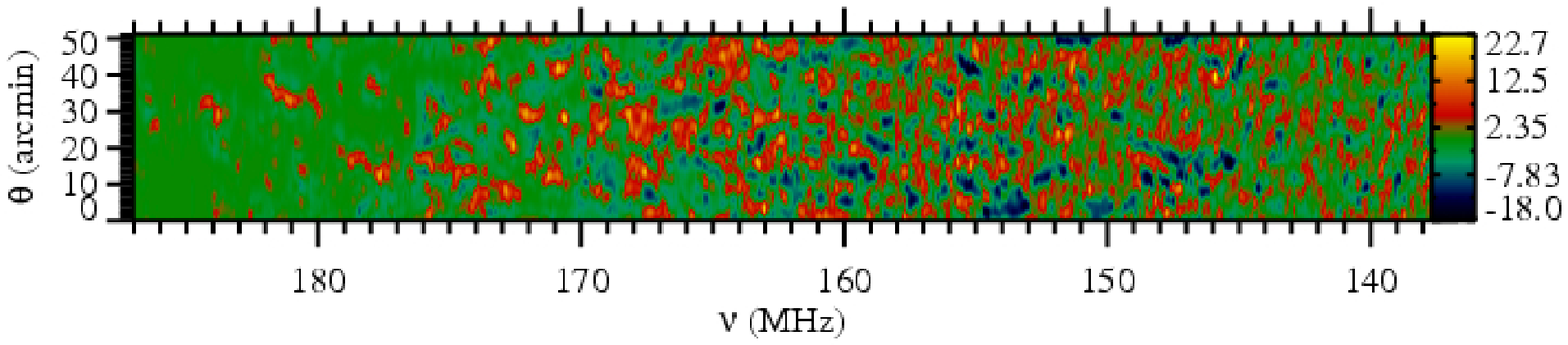}
\vspace{-0.3cm}
\caption{
  Large-scale geometry of reionization and the large local variations in
  reionization history as seen at redshifted 21-cm line: (top) $\lg(\delta
  T_b)$ and (bottom)  $\delta T_b$ smoothed with 3' compensated Gaussian beam.
\label{pencil}}
  \includegraphics[width=3.3in]{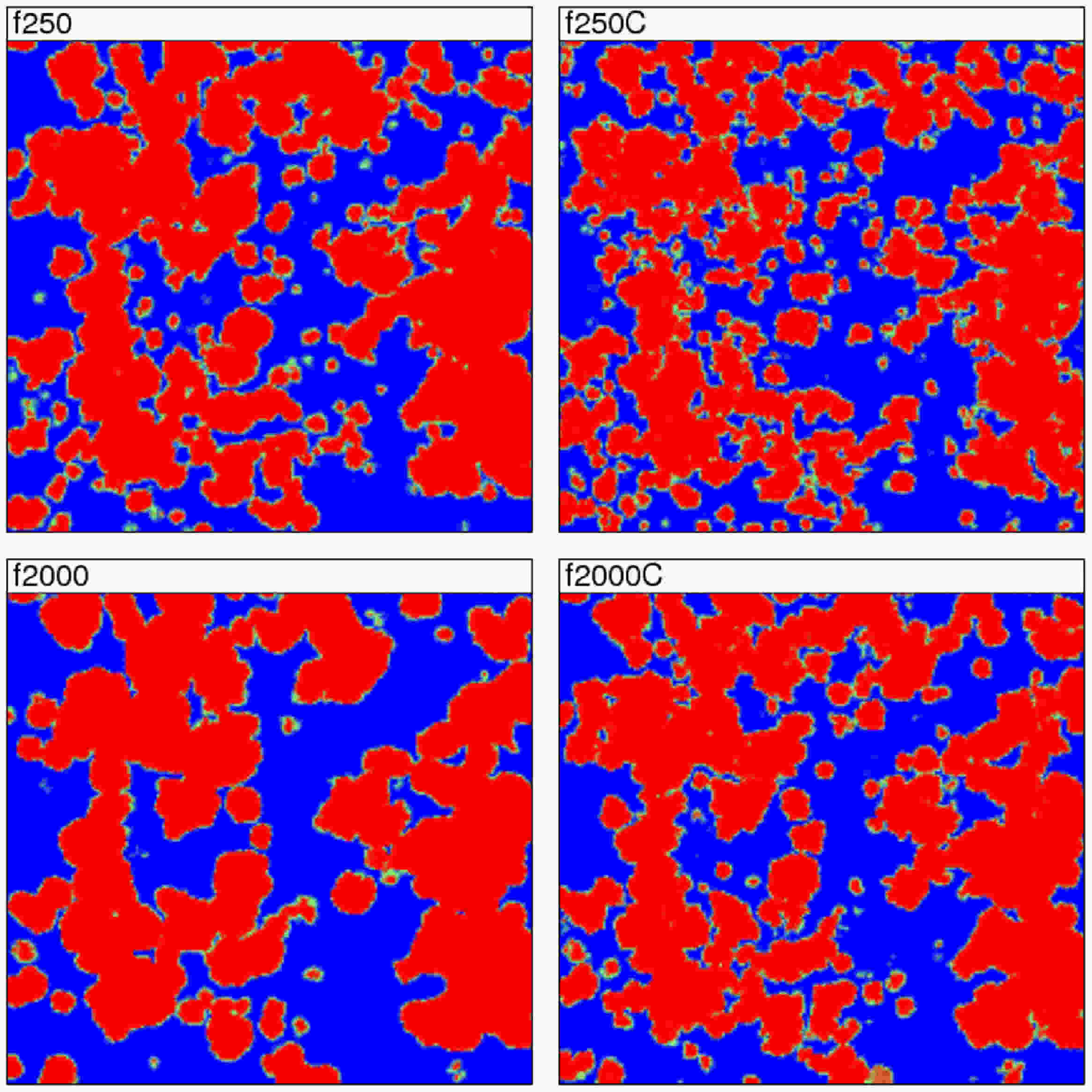}
\caption{Ionization maps (blue neutral, red ionized) of selected
  $100\ h^{-1}$Mpc box simulations for high (bottom panels) and low source
  efficiencies (top) and with (right panels) and without (left panels)
  sub-grid clumping.
\label{scales}}
\end{figure}

\begin{figure}
  \includegraphics[width=3in]{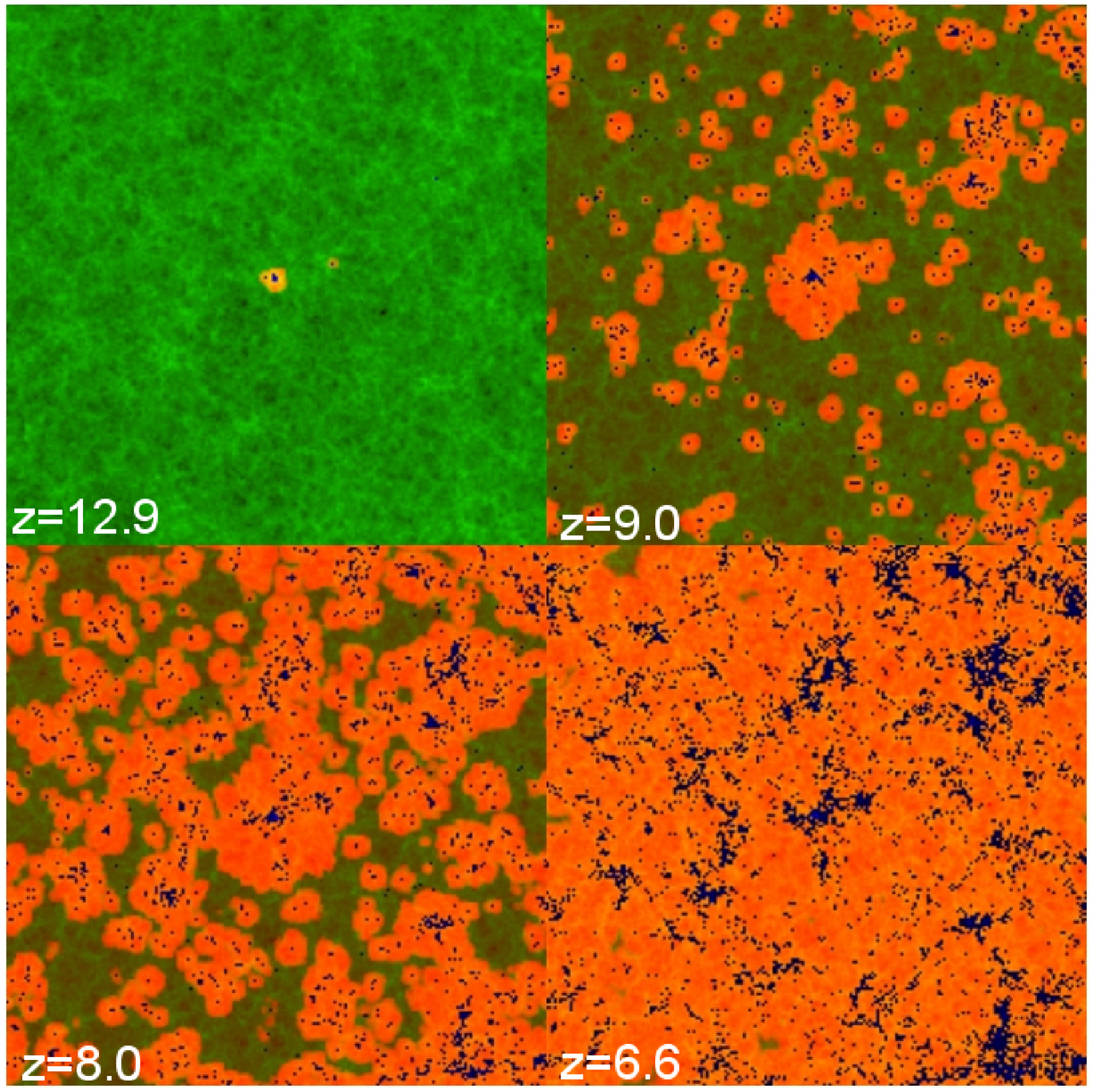}
\caption{The reionization history of a high density peak.
The images are centered on the most massive (at $z=6$) halo
in our computational volume and are of size $100\,h^{-1}$Mpc
to the side. The snapshots are (left to right): $z=12.9$,
$z=9.0$, $z=8.0$, and $z=6.6$. The underlying cosmological 
density field (dark,green) is superimposed with the ionized 
fraction (light, orange) and the ionizing sources (dark, blue  
dots).
\label{peak_evol}}
  \includegraphics[width=3.3in]{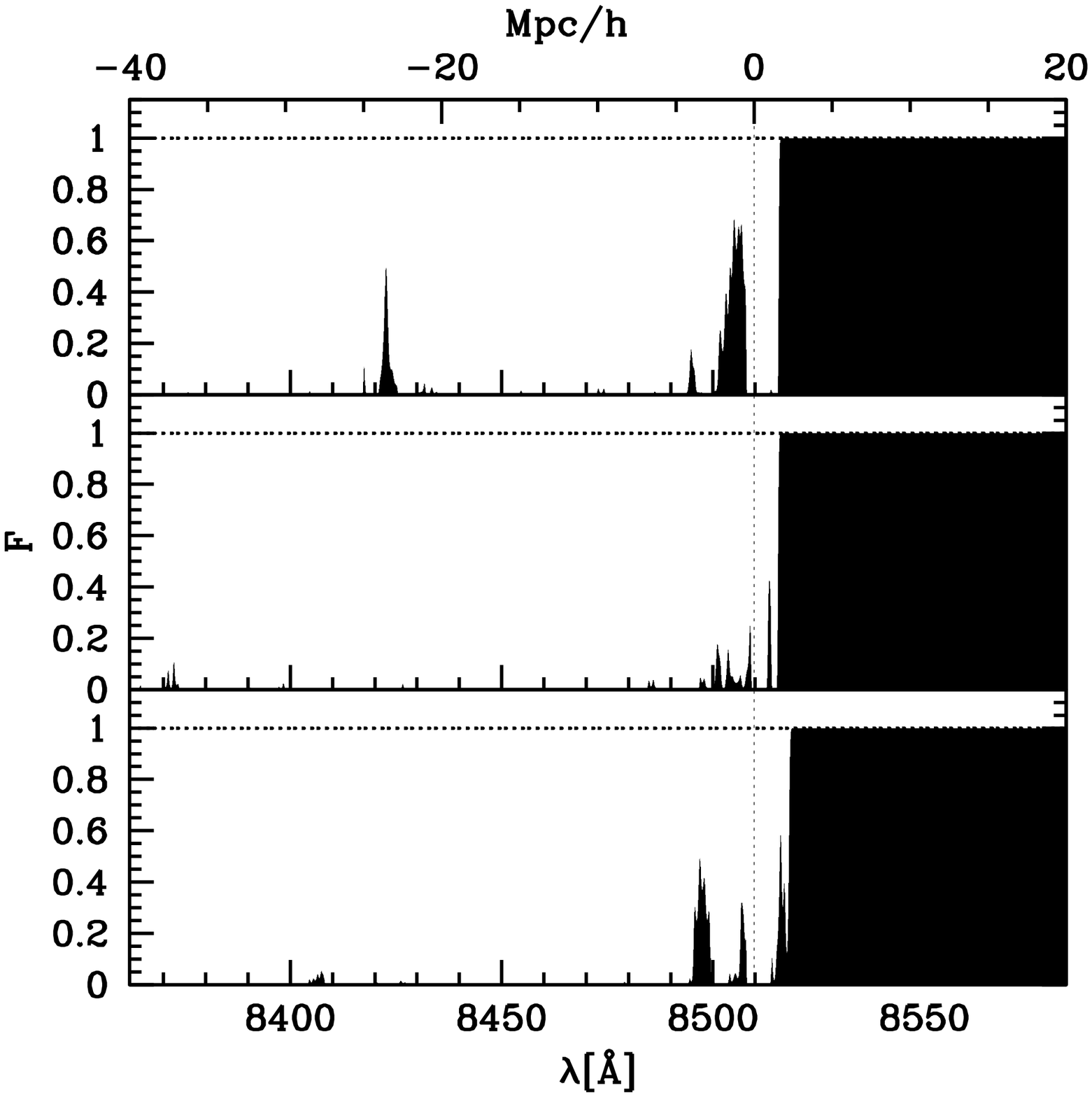}
\caption{Ly-$\alpha$ sources at redshift $z=6.0$: Transmission factor along 
sample lines-of-sight vs. $\lambda/$comoving distance from the most massive 
galaxy in the computational volume. There is significant transmission in the
proximity zone and transmission gaps in the mean IGM away from the
source. Gas infall results in some absorption red-ward of the line center.
\label{spectra2}} 
\end{figure}

\section{Introduction} 
The observations of high-redshift QSO's
\citep{2001AJ....122.2833F,2001AJ....122.2850B} and large-scale CMB
polarization \citep{2007ApJS..170..377S} indicate that the intergalactic
medium has been completely ionized by redshift $z\sim6$ through an extended
process. The most probable cause was the ionizing radiation of the First Stars
and QSO's. Currently these are the two main direct observational constraints
on this epoch. This scarcity of observational data is set to change
dramatically in the next few years, however. A number of large observational
projects are currently under way, e.g. observations at the redshifted 21-cm
line of hydrogen \citep[e.g.][]{1997ApJ...475..429M,2000ApJ...528..597T, 
2002ApJ...572L.123I,2006MNRAS.372..679M,2006PhR...433..181F}, detection of 
small-scale CMB anisotropies due to the kinetic Sunyaev-Zel'dovich (kSZ) 
effect \citep[e.g.][]{2000ApJ...529...12H,2003ApJ...598..756S,kSZ}, and 
surveys of high-redshift Ly-$\alpha$ emitters and studies of the IGM
absorption
\citep[e.g.][]{2003AJ....125.1006R,2004ApJ...604L..13S,
2006NewAR..50...94B}. 
The planning and success of these experiments relies critically upon 
understanding the large-scale geometry of reionization, i.e. the size- and 
spatial distribution of the ionized and neutral patches. This is best 
derived by large-scale simulations, although a number of semi-analytical
models exist as well \citep[e.g.][]{2004ApJ...613....1F}. Recently we presented
the first large-scale, high-resolution radiative transfer simulations of cosmic 
reionization
\citep{2006MNRAS.369.1625I,
2007MNRAS.376..534I} and applied those to derive a range of reionization
observables \citep{2006MNRAS.372..679M,kSZ,pol21,cmbpol,wmap3,2007arXiv0708.3846I}.
Here we summarize recent results on the characteristic scales and topology of
reionization and implications of our simulations for the observability of
high-redshift Ly-$\alpha$ sources.

\section{Characteristic scales of Reionization}
\begin{figure}[ht]
  \includegraphics[width=3.in]{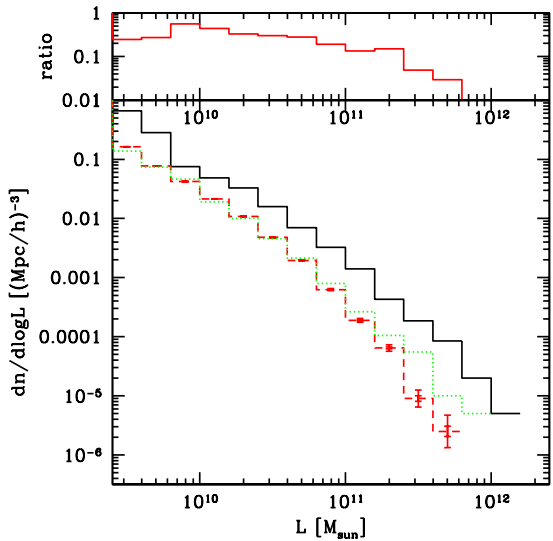}
\vspace{-5mm}
\caption{Ly-$\alpha$ sources at redshift $z=6.0$: Luminosity function without 
(black) and with absorption included (red). For reference, the green, dotted 
line shows the result suppressing each source by 50\%, which would be the case
if e.g. all of the blue wing of the emission line, but none of the red wing is
absorbed.
\vspace{-5mm} 
\label{lum}}
\end{figure}

The characteristic scales of the ionization and density fluctuations are
directly imprinted in the fluctuations of all reionization observables 
(Fig.~\ref{pencil}). It is thus critical to understand what these scales are
and their dependences on the (still not yet well-known) reionization parameters.
We investigated the dependence of the characteristic sizes of ionized and
neutral patches on some basic properties of the sources of reionization and
the intergalactic medium, such as the halo mass-to-light ratio, susceptibility
of haloes to positive and negative feedback, and the gas clumping at small
scales. We used two independent methods for identifying the size distribution
of patches, the friends-of-friends (FOF) method \citep{2006MNRAS.369.1625I},
where all topologically-connected ionized regions are considered as one region,
and the spherical average method \citep{2007ApJ...654...12Z}, where an
averaging over a spherical region and an ionization threshold are used to
define the size. In the FOF method, throughout most of the evolution there is
one very large connected region in which most of the volume is contained. For
the spherical average method, the bubble distribution typically peaks on Mpc
scales. Suppression of ionizing sources within ionized regions and gas
clumping both reduce the size and increase the number of H~II regions, 
although the effect is modest, reducing the typical radius of H~II regions 
by factors of at most a few (Fig.~\ref{scales}). We also found that density and ionized 
fraction are correlated on large scales, regardless of the degree of 
clumping and suppression. The genus of the ionization field proved to 
be much more sensitive to suppression and clumping than the size 
distributions or the power spectra and could thus be used for more 
detailed characterization of the reionization parameters. We will present
these results in detail in an upcoming paper (Alvarez et al., in prep.).

\section{High-redshift Ly-$\alpha$ sources}
Observations of the high-redshift Ly-$\alpha$ sources have provided us with a
wealth of information about the state of the IGM and the nature of the
galaxies at the end of reionization and hold even more promise for the
future. In addition to probing the IGM and the source luminosity function (and
thus, indirectly, the halo mass function and ionization source properties),
they may also be used to constrain the reionization topology
\citep{2007arXiv0708.2909R}. The first rare objects form in the highest peaks
of the density field. The statistics of Gaussian fields predicts that such peaks
are strongly clustered at high redshift. Hence each high-redshift,
massive galaxy was surrounded by numerous smaller ionizing sources. The
self-consistent simulations of such regions require following a large
volume while also resolving the low-mass halos driving reionization. 

In Figure~\ref{peak_evol} we illustrate several stages of the reionization 
history of a high density peak. The most massive source in our volume (at
$z=6$) is shifted to the centre using the periodicity of the computational 
box. At redshift $z=12.9$ the source is invisible due to the damping wing from 
the neutral gas outside the small H~II region. By redshift $z=9$ many more
haloes have formed, most of them in large clustered groups. The H~II region
surrounding the central peak is among the largest, but the central source
emission remains strongly affected by damping. Only by redshift $z=8$
the ionized region is large enough to render the source potentially
visible. The reionization geometry becomes quite complex, most ionized bubbles
are interconnected, but large neutral patches remain between them, which is
also showing in the topological characteristics of the filed like the genus
discussed above. Finally, by the nominal overlap $z=6.6$ all ionized regions
have merged into one topologically-connected region, although substantial
neutral patches remain interspersed throughout our volume, which is still
largely optically-thick to both Ly-$\alpha$ and ionizing radiation. Only by
$z\sim6$ this volume becomes on average optically-thin to ionizing radiation.

In Figure~\ref{spectra2} we show some sample absorption spectra for the
same luminous source. The spectra exhibit extended high-transmission (10-60\%
transmission) regions in the highly-ionized proximity zone of the luminous
source, within 5~Mpc$\,h^{-1}$ ($\sim20$~\AA). The center of the peak itself
is optically-thick due to its high density. The infall around the central peak
blue-shifts photons, resulting in some transmission behind the redshift-space
position of the source. Away from the proximity region the absorption is
largely saturated, but there are a number of transmission gaps with up to a
few per cent transmission. In Figure~\ref{lum} we show the
Ly-$\alpha$ source luminosity function at the same redshift. For the weaker 
sources  roughly half of the intrinsic luminosity is transmitted (the red wing
of the line), while the most luminous sources suffer from additional
absorption due to the gas infall that surrounds them. Future work will
quantify the statistics of these features and its evolution. 


\bibliographystyle{Spr-mp-nameyear}   


\end{document}